\documentclass[pre,amsmath]{revtex4}

\usepackage{graphicx}
\usepackage{subfigure,dcolumn}
\usepackage[T2A,T1]{fontenc}
\usepackage[russian,english]{babel}
\graphicspath{{figure/}}

\usepackage{listings}
\begin{document}

\title{Phase diagram of the repulsive Blume-Emery-Griffiths model in the presence of external magnetic field on a complete graph}

\author{Soheli Mukherjee$^{1, 2}$}
\email{soheli.mukherjee@niser.ac.in}
\author{Raj Kumar Sadhu$^{3}$}
\email{raj-kumar.sadhu@weizmann.ac.il}
\author{Sumedha$^{1, 2}$}
\email{sumedha@niser.ac.in}
\affiliation{$^1$ School of Physical Sciences, National Institute of Science Education and Research, Jatni - 752050,India}
\affiliation{$^2$ Homi Bhabha National Institute, Training School Complex, Anushakti Nagar, Mumbai 400094, India}
\affiliation{$^3$ Department of Chemical and Biological Physics, Weizmann Institute of Science, Rehovot 7610001, Israel}

\begin{abstract}
  For the repulsive Blume-Emery-Griffiths model the phase diagram in the space of three fields, temperature ($T$), crystal field ($\Delta$), and magnetic field ($H$), is computed on a complete graph, in the canonical and microcanonical ensembles. For weak strength of the biquadratic interaction ($K$), there exists a tricritical point in the phase diagram where three critical lines meet. As $K$ decreases below a threshold value(which is ensemble dependent), new multicritical points like the critical end 
  point and bicritical end point arise in the $(T,\Delta)$ 
 plane. For $K>-1$, we observe that the two critical lines in the $H$ plane and the multicritical points are different in the two ensembles. At $K=-1$, the two critical lines in the $H$ plane disappear and  as $K$ decreases further, there is no phase transition in the $H$ plane.  Exactly at $K=-1$ the two ensembles become equivalent. Beyond that for  all $K<-1$, there are no multicritical points and there is no ensemble inequivalence in the phase diagram. We also study the transition lines in the $H$ plane                                                            for positive $K$ i.e. for attractive biquadratic interaction. We find that the transition lines in  the $H$ plane are not monotonic in temperature for large  positive $K$.
\end{abstract}

\maketitle

\section{Introduction}

The Blume-Emery-Griffiths (BEG) model is the simplest model which incorporates biquadratic interactions\cite{blume1971ising}. Presence of biquadratic exchange interaction is known to be relevant to understand the properties of the rare-earth compounds. The biquadratic exchange was first suggested by Kittel in the theory of magnetoelastic effect in NiAs type structures\cite{kittel1960}, and by Anderson in the superexchange interaction of iron group oxides and flurides\cite{anderson1991}. In rare-earth compounds, the unpaired $4 f$ electrons lie deep inside the $5 d$ and $5 s$ orbital. So these electrons do not experience the strong crystal field generated by other ions in the crystal. Hence their spherically symmetric potential is not completely destroyed.  As a result the orbital angular momentum is not entirely quenched. The super-exchange between these unquenched orbital momentum gives rise to a biquadratic exchange interaction term in the Hamiltonian\cite{birgeneau1969}. Other interactions such as phonon exchange between ions\cite{https://doi.org/10.1002/pssb.2221970127} and the Schrodinger's spin-one exchange operator\cite{allan1967spin} can also result in the inclusion of such interaction. Both attractive and repulsive biquadratic interactions are of interest. The requirement of small repulsive exchange interaction in a Hamiltonian was first mentioned by Harris and Owen \cite{harris1963biquadratic} and Rodbell \textit{et.al} \cite{rodbell1963biquadratic} in order to explain the paramagnetic resonance of the Mn ion pairs which are present as an impurity in the crystals of MgO. 

Biquadratic exchange interaction is represented by a term that is fourth order in spin operators. Spin-1 BEG model has been shown to successfully capture the physics of these higher order interactions and has been widely studied. It incorporates an uniform crystal field($\Delta$) and a biquadratic exchange interaction ($K$) along with the bilinear exchange interaction term.  This model was first introduced in order to explain the phase separation and superfluidity of $^3 He- ^4 He$ mixture \cite{blume1971ising}. Apart from this many other physical systems like: metamagnets, liquid crystals, semiconducting alloys, microemulsions, etc can also be mapped to the BEG model. This model has a rich phase diagram depending on the sign and magnitude of the biquadratic term. The special case, $K=0$ is known as the Blume-Capel model. Blume-Capel model was first studied by M.Blume\cite{blume1966theory} and H.W.Capel \cite{CAPEL1967295} in order to explain the first order transition in $UO_2$. The another extreme case with the zero  bilinear exchange was studied by Griffiths \cite{GRIFFITHS1967689}.

The attractive biquadratic exchange interaction($K>0$) BEG model has been extensively studied. Its phase diagram changes with the value of $K$. For small $K$, there is a transition from a ferromagnetic to paramagnetic phase in the ($T-\Delta$) plane. This transition line changes from a continuous to a first order transition line at a tricritical point(TCP). As $K$ increases further, another paramagnetic state emerges and the two paramagnetic states are separated by another first order line. The two first order lines meet at a triple point. For larger value of $K$, the continuous transition line terminates on the first order line at a critical end point(CEP), and the TCP disappears. The phase diagram has been  well studied using various techniques like mean-field\cite{blume1971ising, hovhannisyan2017complete,PhysRevA.11.2079, mukamel1974ising, PhysRevB.15.441}, cluster variation\cite{buzano1993surface}, Bethe lattice\cite{chakraborty1986spin}, high-temperature series expansion \cite{saul1974tricritical} etc. Apart from the mean-field, the phase diagram has also been studied in the finite dimensions using renormalization group\cite{berker1976blume, BAKCHICH1992524}, Monte-Carlo simulations in two and three dimensions\cite{tanaka1985spin, PhysRevE.70.046111, zierenberg2017scaling}. However, the simulations have been done mostly on the Blume-Capel model to study the continuous transition line and the TCP. The other multicritical points have not been studied as they are hard to locate in the simulations.

For repulsive biquadratic interaction ($K<0$) the competition between the biquadratic and the bilinear interactions gives rise to a very different behaviour from the behaviour for positive $K$. The negative $K$ term chooses the non-magnetic spins over the 
magnetic ones. This creates a competition between the magnetic and non-magnetic spins. 
In a recent study on a complete graph\cite{prasad2019ensemble} in the $(T-\Delta)$ plane it was shown that as $K$ becomes more negative, the TCP changes to a quadrupolar point at $K=-0.0828$ and $K=-0.1838$ in microcanonical and canonical ensembles respectively. They studied the system for small negative values of $K$ upto $K=-0.4$. In this paper, we study the BEG model on a complete graph for the entire range of $K$, with emphasis on large negative $K$ regime in the ($T-\Delta-H$) space. Since the multicritical points occur in systems described by three or more thermodynamic fields, it is useful to study them in ($T-\Delta-H$) space. Though less studied than the attractive BEG model, the model has been studied in the canonical ensemble on bipartite lattices in the past using mean field\cite{hoston1991multicritical,prasad2019ensemble}, renormalization group \cite{hoston1991dimensionality, PhysRevB.47.15019, BRANCO1996477}, Monte Carlo simulations in two and three dimensions\cite{wang1987phase, wang1987phase1, ekiz2002multicritical,rachadi2004monte, netz1992new}, cluster-variation methods\cite{rosengren1993phase}, and hierarchical models\cite{doi:10.1063/1.358047} .

\begin{figure}
\centering
\includegraphics[scale=0.4]{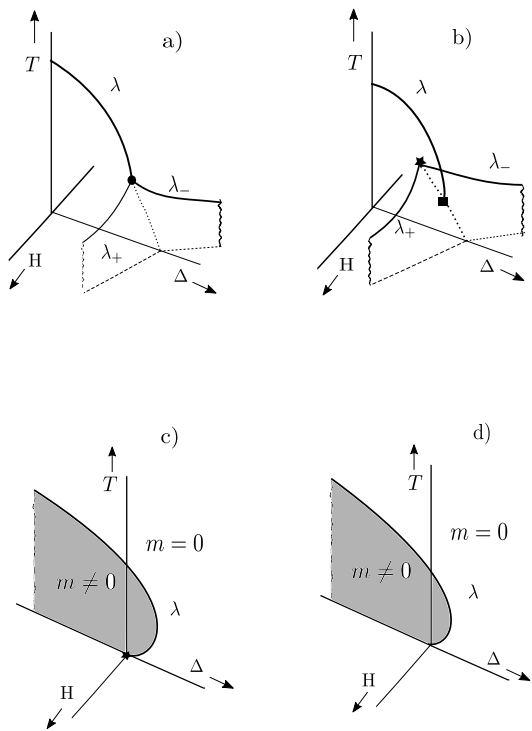}
\caption{Schematic phase diagram of the repulsive BEG model in the ($T-\Delta-H$) space for both canonical and microcanonical ensembles. Solid lines represent the critical lines($\lambda$, $\lambda_\pm$) and the dashed lines represent the lines of first order transition. The $\lambda$ line is the line of continuous transition between the ferromagnetic phase($m \neq 0$) and the paramagnetic phase($m=0$) in the $H=0$ plane, whereas the $\lambda _{\pm}$ lines  are the line of continuous transition in the $\pm H $ planes respectively. The solid circle represents the tricritical point(TCP), where the $\lambda$ and $\lambda_\pm$ lines meet. The star symbol represents the bicritical end point(BEP), where the $\lambda_\pm$ lines meet inside the ordered region. The square symbol represents the critical end point(CEP), where the $\lambda$ line terminates on the first order line. 
\textbf{(a)} Shows the phase topology in the \textit{Canonical ensemble}, for the range: $-0.1838 \leq K \leq 0$ and in the, \textit{Microcanonical ensemble} for: $-0.0828 \leq K \leq 0$. In this regime the critical lines($\lambda _{\pm}$) meet the $\lambda$ line at the TCP.
 \textbf{(b)} Is the phase topology in the \textit{Canonical ensemble} for the range: $-1 < K < -0.1838$ and, in the \textit{Microcanonical ensemble} for the range: $-1 < K < -0.0828$. Here the $\lambda _{\pm}$ lines move inside the ordered region and meet at the BEP. The $\lambda$ line terminates on the first order line at a CEP. 
 \textbf{(c)} Shows the phase topology for both \textit{Canonical} and, \textit{Microcanonical ensembles} at $K=-1$. In both the ensemble the wings as well as the BEP and CEP reaches $T=\Delta=0$.
  \textbf{(d)} Topology of the phase diagram for $K < -1$ for both the \textit{Canonical} and, \textit{Microcanonical ensembles}. Only the $\lambda$ transition remains. The only transition is from the ferromagnetic state to the paramagnetic state in the $H=0$ plane. }
\label{k<0}
\end{figure}
We solve the phase diagram both in the microcanonical and canonical ensemble. We observe four topologies of the phase diagram depending on the values of $K$. We find that for low negative values of $K$ there is a  TCP at which three critical lines($\lambda, \lambda _{\pm}$) meet(see Fig.\ref{k<0}(a)). The $\lambda$ line is the line of continuous transition between the ferromagnetic phase(with magnetization $m \neq 0$) and the paramagnetic phase(with magnetization $m = 0$) in the $H=0$ plane. The  $\lambda _{\pm}$ lines are the lines of continuous transition in the $\pm H $ planes respectively. The $\lambda _{\pm}$ lines enclose two first order surfaces which meet along a triple line in the $H=0$ plane. Since these first order surfaces appear  symmetrically in the phase diagram like the wings of a bird, they are referred as 'wings'\cite{blume1971ising}. As we decrease $K$, TCP becomes a quadrupolar point at $K=-0.0828$ and $K=-0.1838$ in the microcanonical and canonical ensembles respectively. On reducing $K$ further, the $\lambda _{\pm}$ lines move inside the ordered region and meet at a new multicritical point, the  bicritical end point(BEP), and the $\lambda$ line truncates on the first order line at a CEP as shown in Fig.\ref{k<0}(b). Earlier this BEP was reported as an ordered critical point \cite{hoston1991multicritical, hoston1991dimensionality, PhysRevB.47.15019, BRANCO1996477, wang1987phase, wang1987phase1, ekiz2002multicritical, rachadi2004monte, netz1992new, rosengren1993phase, prasad2019ensemble}. By introducing the external field we realize that this point is a junction of two critical lines($\lambda _{\pm}$) and hence a BEP. We find that the width of the wings in temperature shrinks as $K$ approaches $K=-1$. At exactly $K=-1$ the BEP as well as the CEP moves to $T=\Delta=0$ and the wings vanish(see Fig.\ref{k<0}(c)). On further reducing $K$, we see no transition in the finite $H$ plane. There is only a transition from a ferromagnetic state to a paramagnetic state in the $H=0$ plane(Fig.\ref{k<0}(d)). The area under the $\lambda$ line shrinks as $K$ becomes more and more negative. At $K \longrightarrow - \infty$, only the paramagnetic state survives and there is no transition.

The competition introduced by the repulsive biquadratic interaction makes it a very interesting model to study. In fact we find that 
the phase diagram for the repulsive($-1 < K \leq 0$) BEG model is  similar to the topology of the phase diagram for the Blume-Capel model with random crystal field studied recently\cite{mukherjee2020emergence}  for the intermediate and weak disorder. In this paper, we have looked at the ensemble inequivalence not just by looking at the first order line in the $(T-\Delta)$ plane but also by computing the critical lines($\lambda_{\pm}$) in the $H \neq 0$ plane. We find that these two critical lines are different in the two ensembles in general besides the multicritical points. Another interesting observation we have is that for $K \le -1$ the two ensembles are equivalent. Attractive BEG model has been well studied and we find results similar to as reported in the earlier studies in the $(T-\Delta)$ plane \cite{blume1971ising, hovhannisyan2017complete,PhysRevA.11.2079, mukamel1974ising, PhysRevB.15.441, berker1976blume, BAKCHICH1992524, tanaka1985spin, buzano1993surface, chakraborty1986spin, saul1974tricritical}. But in the $(T-\Delta-H)$ space, we found a  non-monotonic behaviour of the wings in terms of temperature as $K$ becomes greater than $K=1$. This as far as we know has not been reported earlier. 

The plan of the paper is as follows: In Section \ref{sec2} we introduce the BEG model and discuss its zero temperature phase diagram. In Section \ref{sec3} and Section \ref{sec4} we derive the equations of the critical lines in the ($T-\Delta-H$) space for both repulsive and attractive BEG model for the canonical ensemble and the  microcanonical ensemble respectively. In Section \ref{sec5} we  discuss the ensemble inequivalence in detail. We conclude in Section\ref{sec6}.


\section{Model}\label{sec2}

The Hamiltonian of the BEG model on a complete graph in the presence of external magnetic field is given by:

\begin{equation} \label{eq:h}
\mathcal{H}=- \frac{1}{2N} (\sum_{i} S_i)^2- \frac{K}{2N} (\sum_{i} S_i^2)^2 +\bigtriangleup \sum_{i} S_i^2-H\sum_{i} S_i
\end{equation}
where $S_i$ can take three values $\pm1$, 0, $H$ is a constant external field coupled with the order parameter, $\bigtriangleup$ is the crystal field, and $K$ is the biquadratic interaction coefficient. The two order parameters are: magnetization, $x_1=\sum_i \frac{S_i}{N}$ and the density of the $\pm 1$ spins, $x_2=\sum_i \frac{S_i^2}{N}$. For any finite $K$, as $\Delta \rightarrow - \infty$, this model becomes equivalent to the Ising model, as the spins take only $\pm 1$ values. As $\Delta$ increases, the number of vacancies increases in the system. For negative $K$, spins are more likely to take value $0$. At finite temperature, when $K <0$, both the biquadratic term and the crystal field term prefer $0$ spins. Hence the $\lambda$ transition occurs at a lower $\Delta$ as $K$ decreases. On the other hand, for positive $K$ the magnetic spins are more likely to be chosen. Hence, when $K > 0$ there is a competition between the biquadratic and crystal field term in the Hamiltonian and the $\lambda$ transition occurs at a higher $\Delta$ for positive $K$.

First let us look  at the zero temperature phase diagram of the system. The energy per particle can be written as(from the Hamiltonian Eq.(\ref{eq:h})): $\epsilon=- \frac{1}{2} (x_1^2 +K x_2^2)+ \Delta x_2 -H x_1 $. When all the spins are zero the energy is $\epsilon=0$. Apart from this paramagnetic phase, there are other states which are possible depending on the parameter values. For $-1 \leq K \leq \infty$, the ferromagnetic state, $x_1=\pm 1$ and $x_2=1$ dominates. Energy of this state is $\epsilon=-\frac{1}{2} (1+K) + \Delta$. If $2 \Delta > 1+K$, then the phase is paramagnetic, for $2 \Delta < 1+K$, the phase is ferromagnetic. At exactly $2 \Delta = 1+K$ there is a first order phase transition. For $K< -1$, the term $-\frac{1}{2} (1+K)$ in the energy contributes a positive value. So for any $\Delta \geq 0$, paramagnetic phase is the stable state. As $\Delta$ becomes negative, there is another ferromagnetic state with $|x_1|=x_2<1$, which becomes stable when $|\Delta| < - \frac{1}{2} (1+K)$. For $|\Delta| > - \frac{1}{2} (1+K)$, the state with $|x_1|=x_2=1$ becomes stable and there is a first order transition now between these two ferromagnetic state at $\Delta=\frac{1+K}{2}$.


\section{Canonical ensemble}\label{sec3}

Given the Hamiltonian(Eq.(\ref{eq:h})), the probability of the spin configuration($C_N= \lbrace S_i \rbrace $) for $N$ spins can be expressed as:
\begin{equation}
P(C_N)= \frac{e^{\beta (\frac{1}{2N} (\sum_{i} S_i)^2+ \frac{K}{2N} (\sum_{i} S_i^2)^2 -\bigtriangleup \sum_{i} S_i^2+H\sum_{i} S_i)}}{Z_N}
\end{equation}
Where $Z_N$ is the partition function and $\beta=\frac{1}{T}$. The free energy of the mean field models can be calculated in many ways. We calculated the free energy of the system using the large deviation principle(LDP)\cite{TOUCHETTE20091}, which states that the probability of a spin configuration having magnetization $x_1$ and density $x_2$ for $N \rightarrow \infty$ can be expressed as:
\begin{equation}
P(C_N) \asymp e^{- N I(x_1, x_2)}
\end{equation}
where $I(x_1, x_2)$ is called the rate function. For a system to satisfy LDP, the following limit should hold:
\begin{equation}
I(x_1, x_2) \asymp - \lim_{N \rightarrow \infty} \frac{1}{N} \ln{P(C_N)}
\end{equation} 
The rate function $I(x_1, x_2)$ can be seen as the full Landau free energy functional for the system. Minimizing the rate function w.r.t $x_1$ and $x_2$ gives the free energy of the system. The detailed calculation of the rate function $I(x_1, x_2)$ is shown in Appendix\ref{app2}. Minimization of the rate function w.r.t $x_1$ and $x_2$ gives the following two coupled equations for the two order parameters:
\begin{eqnarray}
m =\frac{2 e^{\beta(K q-  \Delta}) \sinh {\beta (m+H)}} {1 + 2 e^{\beta(K q- \Delta)} \cosh {\beta (m+H)} } \label{eq:5}\\
q =\frac{2 e^{\beta(K q-  \Delta}) \cosh {\beta (m+H)}} {1 + 2 e^{\beta(K q- \Delta)} \cosh {\beta (m+H)} } \label{eq:6}
\end{eqnarray}
where $m$ and $q$ are the extremums of $x_1$ and $x_2$. For $m \neq 0$, the two fixed point equations are connected via:
\begin{equation} \label{eq:7}
q=m \coth{\beta (m+H)}
\end{equation}
and the free energy at the fixed point can be written as(putting Eq.(\ref{eq:7}) in Eq.(\ref{freeenergy1}));
\begin{eqnarray}
f(m) &=& \frac{\beta m^{2}}{2} +\frac{\beta K m^2 \coth^2{\beta m}}{2} +\log (1+2 e^{-\beta \Delta} \cosh \beta H) \nonumber \\
&-& \log (1+2 e^{\beta(K m \coth{\beta m}-\Delta)} \cosh \beta (m+H) ) 
\label{freeenergy}
\end{eqnarray}
For $H=0$, the system has a line of continuous transition($\lambda$ line) in the ($T-\Delta$) plane. The equation of this line can be obtained by  linearizing  Eq.(\ref{eq:7}). On linearizing we get the equation of the $\lambda$ line to be:
\begin{equation}
2 (\beta-1)= e^{\beta \Delta -K} 
\label{eq:cont}
\end{equation}
For $H \neq 0$, the system has a line of continuous transition in $H_+$ and $H_-$ planes separating the two magnetic states. To calculate these lines we take  $f'(m)=f''(m)=f'''(m)=0$ and $f''''(m)>0$, to get the locus of the critical points. This gives the following equations:

\begin{eqnarray}
&& f_1 \equiv m - \frac{a z}{C} = 0 \label{eq:cp} \\ 
&& f_2 \equiv \frac{1}{z a \beta} - \frac{a z y^2+\sqrt{y^2+1} y +y K (\sqrt{y^2+1}- \beta m y^2)}{C^2}  = 0 \label{eq:cp2}\\ 
&& f_3 \equiv \frac{\beta^3 z}{C^3}  \Bigg (a z K^2\sqrt[\frac{3}{2}]{y^2+1}-K(y^2+1) y \Big [(2+K)+2 a z (1+K) m y \beta \Big ] \nonumber \\
&&+\sqrt{y^2+1} \Big [a z (1+ 2 K)+ 2 \beta K^2 m y^3+ \beta{^2} a z K^2 m^2 y^4 \Big ]  \nonumber \\
&& -y \Big [1- 2 a^2 z^2+ 2 \beta K a z y m- 2 K y^2- 2 \beta a z K m y^3  + \beta{^2} K^2 m^2 y^4 \Big ] \Bigg)=0 \label{eq:cp3}
\end{eqnarray}

\begin{figure}
\centering
\includegraphics[scale=2.1]{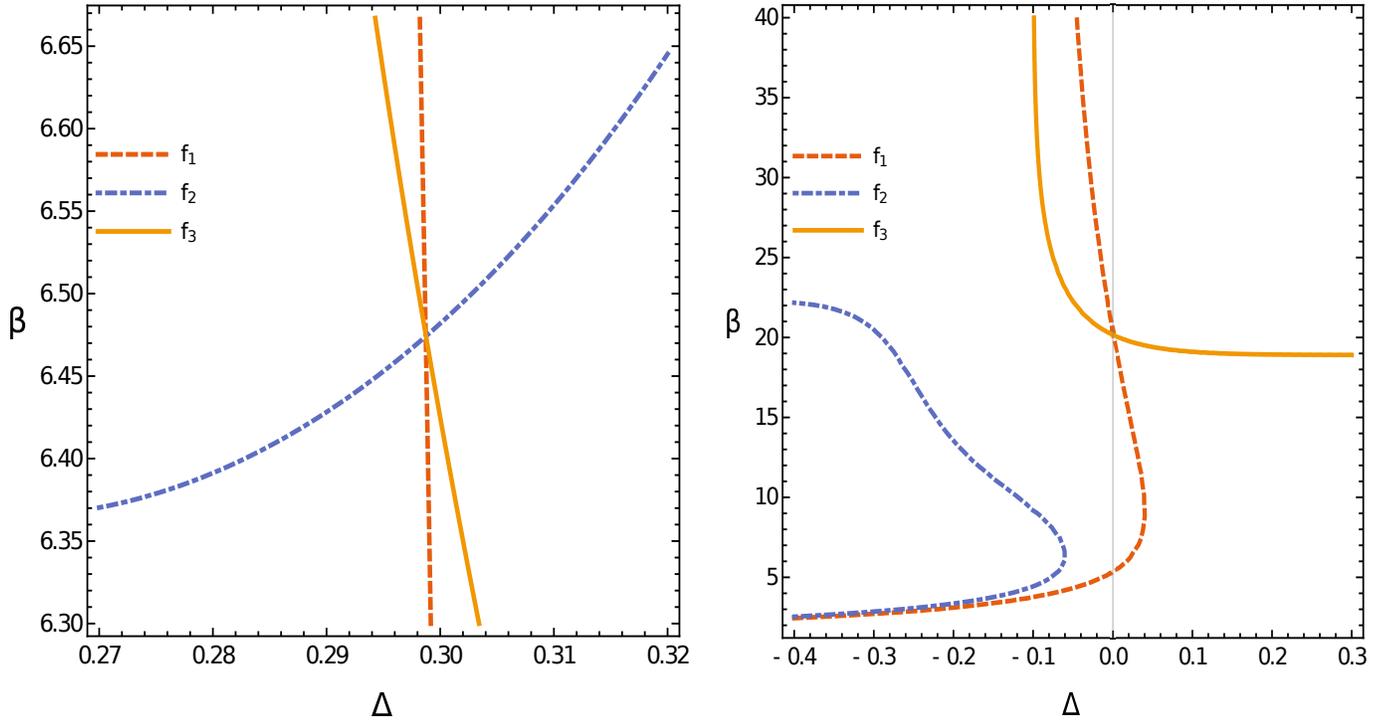}
\caption{Plot of $f_1$, $f_2$ and $f_3$ in $\beta-\Delta$ plane. \textbf{(a)} $K=-0.4$ at $H=0$. The intersection of the three derivative lines give the critical point at a non-zero value of $m$, which gives the locus of the BEP at which the $\lambda_{\pm}$ lines meet in the $H=0$ plane. \textbf{(b)} $K=-2$ at $H=0$. The three lines never intersect simultaneously for any value of $m$, which shows that there are no critical points.}
\label{fig2a7b}
\end{figure}
where $a=e^{- \beta \Delta}$, $y=cosech{\beta (H+m)}$, $z= 2 e^{\beta K m \coth{\beta (H+m)}} $, and $C=\sqrt{y^2+1} z a+y$. For $K=0$, the above equations reduce to the following equations for pure Blume-Capel model:
\begin{eqnarray}
&&  m = \pm \frac{2 a \sqrt{x^2-1}}{(2 a x +1)}  \label{purewings0}\\
&& \frac{2 a+ x }{(2 a x +1)^2} =\frac{1}{2 a \beta} \label{purewings1}\\
&& \frac{ 8 a^2+ 2 a x-1}{(2 a x +1)^3} =0
\label{purewings2}
\end{eqnarray}
with $x= \cosh {\beta(m+H)}$. The solutions of Eq.(\ref{purewings1})-(\ref{purewings2}) are hence given by:
\begin{eqnarray}
x &= & \cosh\beta(m+H)=\frac{\beta -2}{\sqrt{4-\beta}} \\
a &=& e^{- \beta \Delta}=\frac{\sqrt{4-\beta}}{4}
\end{eqnarray}
and the critical lines for $H \neq 0$ plane are the following:
\begin{eqnarray}
m &=&  \pm \sqrt{\frac{\beta -3}{\beta}}\\
H &=& \pm \frac{1}{\beta}\log(\frac{\beta-2+\sqrt{\beta^2-3\beta}}{\sqrt{4-\beta}})- m
\end{eqnarray}
These lines are the $\lambda _{\pm}$ lines\citep{blume1971ising}(depending on the sign of $H$). These lines enclose two first order surfaces in $H \neq 0$ plane called the wings.

For $K \neq 0$ solving Eq.(\ref{eq:cp})-(\ref{eq:cp3}) is not possible analytically. Hence we use graphical methods to get the co-ordinates of the critical points in the ($T-\Delta-H$) space for a given $K$. We plot $f_1$, $f_2$ and $f_3$ in ($\beta-\Delta$) plane for $m$, fixing $K$ and changing different values of $H$. The value of $m$  for a fixed value of $H$ and $K$ at which three equations meet gives the co-ordinates of the critical point. If we now take $H=0$ in Eq.(\ref{eq:cp}), (\ref{eq:cp2}) and (\ref{eq:cp3}), then we will get the co-ordinates of the point of intersection of the $\lambda_{\pm}$ lines in the ($T-\Delta$) plane. We can hence use this to locate the multicritical points(TCP and BEP) in the ($T-\Delta$) plane. We use this to obtain the phase diagram for various values of $K$. For example: Fig.\ref{fig2a7b}(a) is the contour plot of $f_1$, $f_2$ and $f_3$ in the ($\beta-\Delta$) plane at $K=-0.4$ and $H=0$. The intersection of the three functions gives the co-ordinates of the critical point. We find that for $K> -0.1838$ the functions intersect only for $m=0$, which is the point where the $\lambda_{\pm}$ lines meet the $\lambda$ line. Hence this point is the TCP. For the range $-0.1838 > K > -1 $ we find that the intersection occurs for $m \neq 0$. This $m \neq 0 $ solution gives the locus of the BEP where the $\lambda_{\pm}$ lines meet the $H=0$ plane in this regime of $K$. Interestingly, we find that for $K<-1$ the three functions never intersect at the same point for any $m$. For example in Fig.\ref{fig2a7b}(b) we plot three functions for $K=-2$. We will discuss these results in detail in the next section(Sec\ref{rep_beg_can}).


\subsection{Repulsive Blume-Emery-Griffiths model} \label{rep_beg_can}

\begin{figure}
\includegraphics[scale=0.67]{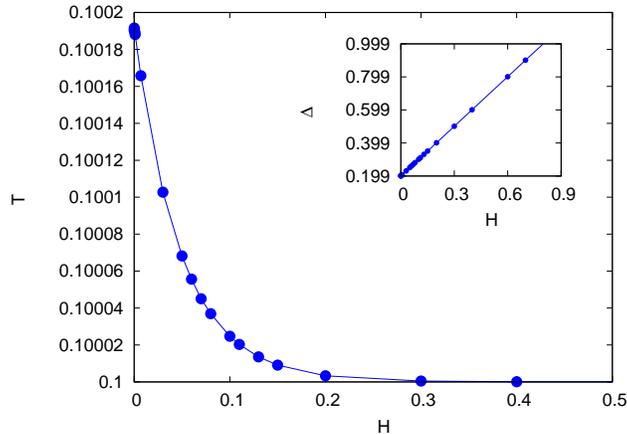}
\caption{The value of temperature($T$) and the crystal field($\Delta$) as a function of $H$ along the $\lambda_+$ line in canonical ensemble for $K=-0.6$. The main plot shows that the temperature decreases exponentially with $H$ and it saturates towards a certain temperature($T_{sat}$) for high magnetic field. The inset shows how $\Delta$ increases linearly with $H$.}
\label{fig:11}
\end{figure}

\begin{figure}
\includegraphics[scale=0.67]{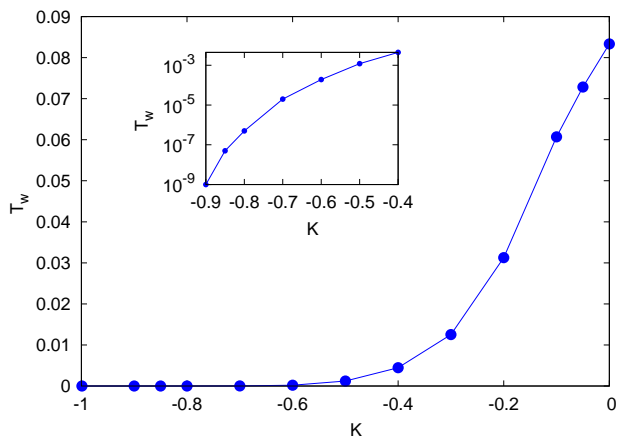}
\caption{The width of the wings in temperature($T_w$) as a function of $K$ for the repulsive BEG model. The main plot shows that as $K$ decreases, the width in temperature goes to zero. The inset is the semi-log plot for the same.}
\label{fig:8}
\end{figure}

In this section we analyze the results of the repulsive BEG model. We find that for $0 \geq  K \geq -0.1838$ two critical lines($\lambda _{\pm}$) at $H \neq 0$ meets the $\lambda$ line(at $H=0$) at the TCP. Temperature  decreases exponentially and $\Delta$ increases linearly with increasing $H$  along the $\lambda_{\pm}$ lines, as shown in Fig.\ref{fig:11}. As $K$ becomes more negative(for $-0.1838 > K > -1$), the ($\lambda _{\pm}$) lines no longer meet the $\lambda $ line, instead they enter into the ordered region and meet at the first order surface($H=0$) at a BEP. The $\lambda$ line in this case terminates on the first order line at a CEP.

As $K$ approaches $-1$, the wing width in temperature(which is the difference between the temperature at the BEP($T_{BEP}$) and the saturation value of the temperature($T_{sat}$) at which both the $\Delta$, $H$ $\rightarrow$ $\infty$) denoted as $T_w$, starts to shrink. At exactly $K=-1$ the BEP, CEP, and the $T_w$ reach zero. As we decrease $K$ further, we find that there are no transitions in the $H$ plane. This is also supported by the fact that now there are no multicritical points in the ($T-\Delta$) plane. Hence we conclude that for $K<-1$ the wing surfaces completely disappear. The phase diagram consists of only a continuous transition line($\lambda$ line) from ferromagnetic phase to paramagnetic phase in the $H=0$ plane. For large negative $K$, the area enclosed by the $\lambda$ line in the ($T-\Delta$) plane  shrinks. At $K \rightarrow - \infty$, there is no phase transition, only the $S=0$ state dominates. The decreasing width of the wings with decreasing $K$ is shown in Fig.\ref{fig:8}. We also observe that the $T_{sat}$ can be approximated numerically as $T_{sat} \simeq (K+1)/4$. This will be discussed in more detail in Sec.\ref{sec:mc_a}, where we obtain the similar results in the microcanonical ensemble. The values of the $T_w$, $T_{sat}$ and the co-ordinates of the multicritical points(TCP, BEP) for the repulsive BEG model are listed in Table(\ref{tab:2}).

\begin{table*}
\begin{center}
\begin{tabular}{|c|c|c|c|c|}
\hline
\multicolumn{5}{|c|}{Canonical: $-1 \leq K \leq 0$}\\
\hline
$K$ & \multicolumn{2}{|c|}{TCP / BEP}  &  $T_{sat}$ &  $T_w$\\
\hline
 & $T$  & $\Delta$ & $\simeq (K+1)/4$ &  (= $T_{TCP/BEP} - T_{sat})$\\
\hline
0 &0.33333 &0.462098 &0.25    &0.0833333   \\
\hline
-0.05 &0.3103448 &0.44741 &0.237501  &0.072843316  \\
\hline
-0.1 &0.2857142 &0.431268 &0.2250124  &0.0607018  \\
\hline
-0.2 &0.2312737 &0.391831 &0.2  &0.03127376  \\
\hline
-0.3 &0.1875335 &0.346377 &0.17499956  &0.01253394  \\
\hline
-0.4 &0.15446222 &0.298727 &0.14999925  &0.00446297 \\
\hline
-0.6 &0.100192068 &0.199958 &0.1   &0.000192068 \\
\hline
-0.7 &0.075009189 &0.149998 &0.07500018   &0.000009009 \\
\hline
-0.8 & $\simeq$ 0.05000025& $\simeq$ 0.1  & $\simeq$ 0.049999875  & $\simeq$ 0.000000375 \\
\hline
-0.9 & $\simeq$ 0.024999969 & $\simeq$ 0.05  &$\simeq$ 0.024999968  &$\simeq$ 0.000000001 \\
\hline
-0.999 &$\simeq$ 0.00025 &$\simeq$ 0.0005 &$\simeq$ 0.00025  &$\simeq$ 0.00  \\
\hline

\end{tabular}
\end{center}
\caption{Co-ordinates of the TCP and BEP for different $K$'s. $T_{sat}$ is the saturation value of the temperature at which both the $\Delta$ and $H \rightarrow \infty$ . $T_w$ is  the width of the wing lines for different $K$}
\label{tab:2}
\end{table*}

\begin{figure}
\begin{tabular}{cc}
 \includegraphics[width=2.95in]{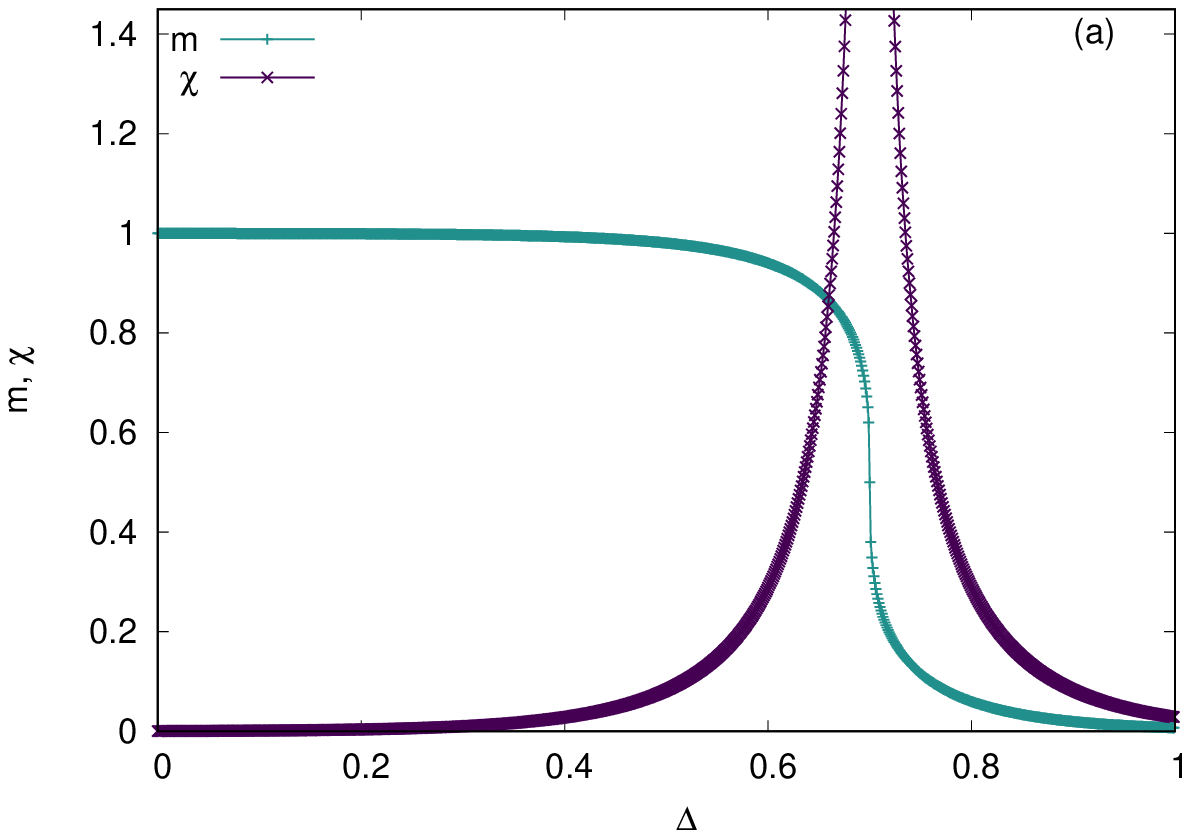}
  \includegraphics[width=3in]{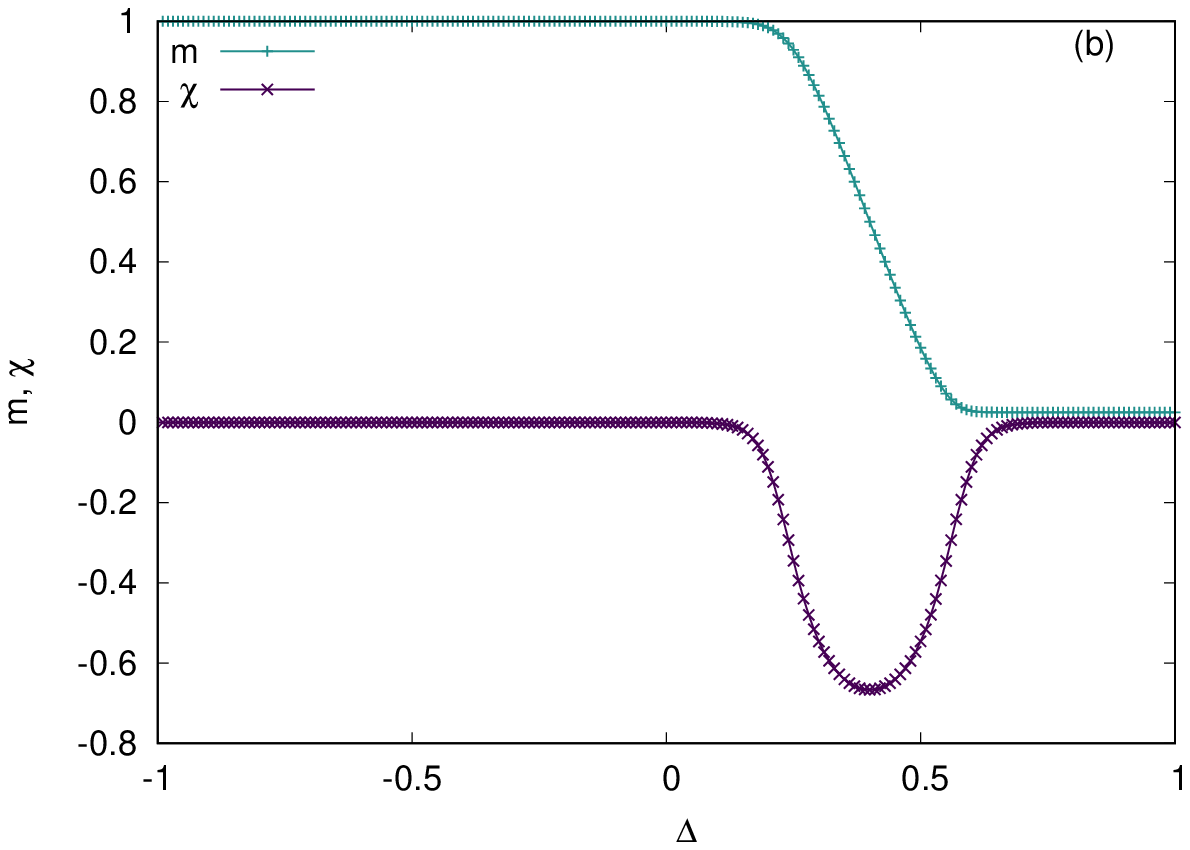}

\end{tabular}
\caption{Magnetization(m) and magnetic susceptibility($\chi$) as a function of $\Delta$ for \textbf{(a)} $K=-0.6$, $T=0.1$ and  $H=0.5$. This shows that the $m$ goes to zero continuously around $\Delta=0.7$. Also the $\chi$ has a singularity at the same $\Delta$ which suggests that there is a second order transition in the $H \neq 0$ plane, \textbf{(b)}  $K=-1.2$ , $T=0.025$ and $H=0.5$. Both the $m$ and $\chi$ changes continuously as a function of $\Delta$. Magnetic susceptibility($\chi$) shows no singularity or discontinuity and there is no phase transition in the finite $H$ plane.}
\label{fig:6}
\end{figure}

Absence of phase transition for $K<-1$ in the $H$ plane can also be seen by looking at the magnetization and susceptibility. We find that the magnetic susceptibility diverges around the expected critical point for $K >-1$. On the other hand for $K \le -1$ magnetic susceptibility is finite in the entire $H$ plane. We plot the magnetization and the susceptibility for  $H=0.5$ at $K=-0.6$ and $K=-1.2$. In Fig.\ref{fig:6}(a) we plot them as a function of $\Delta$ for $K=-0.6$, by fixing $T=0.1$. The susceptibility shows singular behaviour at $\Delta=0.7$. The point of divergence matches with the co-ordinates of the transition obtained from Eq.(\ref{eq:cp}), (\ref{eq:cp2}) and (\ref{eq:cp3}). On the other hand for $K=-1.2$, we find no such divergence. In Fig.\ref{fig:6}(b) for $K=-1.2$ by fixing $T=0.025$ we plot  the magnetization and find that it changes continuously along $\Delta$ and the susceptibility shows a cusp but does not diverge. Though we plot only for a fixed $T$, we have checked the entire plane by changing the values of $T$. Magnetic susceptibility has no divergence for any $T$.

\subsection{Attractive Blume-Emery-Griffiths model } \label{attrac_beg_cano}

The attractive BEG model has been extensively studied earlier by various authors (\cite{blume1971ising}-\cite{mukamel1974ising}) and the topology of the phase diagram is known as a function of $K$ in the ($T-\Delta$) plane. We observe the similar topology of the phase diagram. We study the ($T-\Delta-H$) phase diagram and find that the topology of the phase diagram for different $K$'s are similar to \cite{mukamel1974ising}. To recap we find : For $0 < K \leq 2.78$, the phase diagram is similar to what we find for $0 \geq K \geq -0.1838$.  The $\lambda _{\pm}$ meets at the TCP. For $2.78 < K < 3 $ a new first order surface appears separating two paramagnetic states: P2($m=0$, $q_-<0.5$) and P1($m=0$, $q_+>0.5$). This surface meets the first order line(at $H=0$) at a triple point. This new first order surface terminates on a line of critical points(at $H \neq 0$ plane). As $K$ changes from $K=2.78$, this line of critical points in the paramagnetic region moves higher in temperature and at exactly $K=3$ it intersects the $\lambda _{\pm}$ lines and then extends to infinity. For $3 < K \leq 3.8$, the $\lambda _{\pm}$ lines terminates at the first order surface which separates the P1 and P2 phase, and becomes finite. For $K>3.8$, the $\lambda$ line terminates at a CEP, and thus the wings vanish.

\begin{figure}
\includegraphics[scale=0.67]{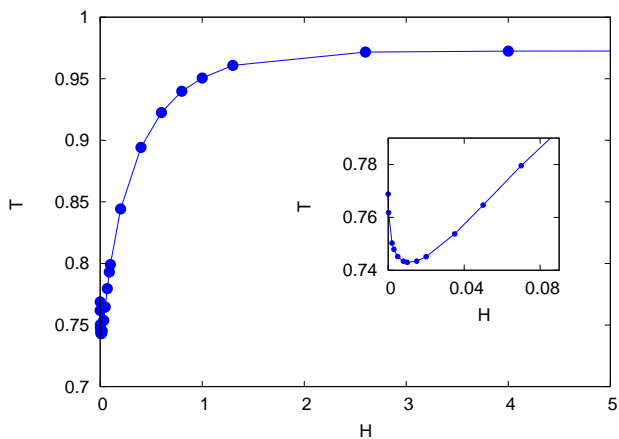}
\caption{Plot for the non-monotonic behaviour of temperature($T$) as a function of magnetic field($H$) along the $\lambda_+$ line for $K=2.89$. The inset shows that for lower $H$, $T$ decreases with $H$ like before, but for higher $H$ it increases and  saturates to a higher value($T_{sat}$) shown in the main plot.}
\label{fig:15}
\end{figure}

We observe that for $K>1$, the wings show non-monotonic behaviour in temperature in contrast to what happens in the range of $-1< K <1$(Fig.\ref{fig:11}). For small values of $H$, the $\lambda _{\pm}$ lines go towards lower $T$ and higher $\Delta$. As $H$ becomes larger, these lines start moving towards higher $\Delta$ and higher $T$ as shown in Fig.\ref{fig:15}. 
This non-monotonic behaviour observed in the wings for all $K \geq 1$ values can be interpreted as follows: For any positive $K$, there are two possible solution of $q$ for a fixed value of energy, $q_{\pm}$(more details are in Sec.\ref{sec4}) with $q_+ > q_-$. When $K \geq 1$, the first term in the energy $\epsilon$(mentioned in Sec.\ref{sec2}) dominates over the crystal field term and the density of $S = \pm 1$ spin increases. For smaller values of $H$ the $q_-$ solution dominates in the system and the wings show monotonic behaviour like before. As $H$ increases further the  $q_ +$ solution becomes favorable which in turn lowers the energy. Thus the energy-entropy balance occurs at a higher temperature.


\section{Microcanonical ensemble}  \label{sec4}
In order to analyze the system in the microcanonical ensemble, we need to express the energy in terms of the number of particles with spin $\pm 1$ and $0$. Let us assume the number of particles with $\pm 1$ spin are $N_\pm$ and the number of particles with zero spins are $N_0$, such that $N = N_+ + N_- + N_0$, where $N$ is the total number of particles in the system. The energy of the system can thus be written as,
\begin{equation}
E=\Delta Q - \frac{1}{2N} M^2 -\frac{K}{2N} Q^2 -H M
\end{equation}
 where $M=N_+ - N_-$ is the total magnetization and $Q=N_+ + N_-$ is the spin density of the system. In terms of $m (=M/N)$ and $q (=Q/N)$, the expression for the energy will be,

\begin{equation}
\epsilon = \Delta q - \frac{1}{2} m^2 - \frac{K}{2} q^2 -H m
\label{eq:energy}
\end{equation}
where, $\epsilon=\frac{E}{ N}$ is the energy per particle, $m$ and $q$ are the single site magnetization and density(as mentioned in Sec.\ref{sec2}). The total number of microstates of the system can be written in terms of $N$, $N_+$, $N_-$ and $N_0$ as,

\begin{equation}
\Omega = \frac{N !}{N_+ ! N_- ! N_0 !}
\end{equation}

In the limit when $N_+$, $N_-$, $N_0$ are large, the expression for entropy, i.e., $S = k_B ~ln (\Omega)$ can be written by using Stirling approximation as, 

\begin{equation}
s= \frac{S}{k_B N} =q~ ln(2) - (1-q) ln (1-q) -\frac{1}{2} (q+m) ln (q+m) - \frac{1}{2} (q-m) ln (q-m) 
\label{eq:entropy}
\end{equation}
where, $s$ is the entropy per particle of the system. The equilibrium entropy can be obtained by maximizing the entropy of Eq.(\ref{eq:entropy}) with respect to $m$ and $q$. We can express $q$ in terms of $m$ and the other variables as,

\begin{equation}
 q_\pm = \frac{\Delta}{K} \pm \gamma ^{1/2}
\end{equation}
where, $\gamma = \Big(\frac{\Delta}{K}\Big)^2 - \frac{2\epsilon}{K} - \frac{m^2}{K} -\frac{2 H m}{K}$. For $K=0$, the expression has a much simpler form, $q= \frac{1}{\Delta} \big( \epsilon+ \frac{1}{2} m^2 + H m\big)$.

Since, there are two values of $q$, the one which is in the range $[0,1]$ will be accepted. There is also a possibility that both the $q$ values are in the range $[0,1]$, then the equilibrium entropy will be the one with maximum value at its corresponding equilibrium $m$. We find that for $K<0$, only $q_-$ is acceptable, however, for $K>0$, both the $q_\pm$ solutions are acceptable.

Next, we aim to find the second order transition line in the ($T- \Delta- H$) space. In the $H=0$ plane, the value of the magnetization $m$ on the line of continuous transition is zero, however, for any nonzero $H$, the magnetization $m$ will have a nonzero value on the continuous transition line. In order to obtain this continuous transition line, we need to equate the first three derivatives of $s$ (with respect to $m$) to zero, with the constraint that the fourth derivative will be negative. The first four derivatives of the entropy $s$ are:

\begin{equation}
\frac{\partial s}{\partial m}= q' ~ ln \Big\{\frac{2(1-q)}{\sqrt{q^2-m^2}}\Big\} - ln \sqrt{\frac{q+m}{q-m}}
\label{ds}
\end{equation}

\begin{equation}
\frac{\partial^2 s}{\partial m^2}= q'' ~ ln \bigg\{\frac{2(1-q)}{\sqrt{q^2-m^2}}\bigg\} - \frac{q'^2}{1-q} - \frac{1}{2} \bigg\{\frac{(q'+1)^2}{q+m} + \frac{(q'-1)^2}{q-m}\bigg\}
\label{dds}
\end{equation} 

\begin{equation}
\frac{\partial^3 s}{\partial m^3}= q''' ~ ln \bigg\{\frac{2(1-q)}{\sqrt{q^2-m^2}}\bigg\} - \frac{q'^3}{(1-q)^2} + \frac{1}{2} \bigg\{\frac{(q'+1)^3}{(q+m)^2} + \frac{(q'-1)^3}{(q-m)^2}\bigg\} - \frac{3}{2} \bigg\{\frac{2 q' q''}{1-q} +\frac{q''(q'+1)}{q+m} + \frac{q''(q'-1)}{q-m}\bigg\}
\label{ddds}
\end{equation}

\begin{eqnarray}
\frac{\partial^4 s}{\partial m^4}= q^{''''} ~ ln \bigg\{\frac{2(1-q)}{\sqrt{q^2-m^2}}\bigg\} -2 q''' \bigg\{\frac{q'+1}{q+m} + \frac{q'-1}{q-m} + \frac{2q'}{1-q}\bigg\} +3 q'' \bigg\{ \frac{(q'+1)^2}{(q+m)^2} + \frac{(q'-1)^2}{(q+m)^2} - \frac{2q'^2}{(1-q)^2} \bigg\} \nonumber \\
- \frac{3}{2} q''^2 \bigg\{ \frac{2}{1-q} + \frac{1}{q+m} + \frac{1}{q-m} \bigg\} - \frac{(q'+1)^4}{(q+m)^3} - \frac{(q'-1)^4}{(q-m)^3} -\frac{2 q'^4}{(1-q)^3}
\label{d4s}
\end{eqnarray}

where, $q'$, $q''$ ...... are partial derivatives of $q$ w.r.t. $m$. We solve the above first three equations numerically and obtain a set of physical solutions ($\Delta$, $\epsilon$, $m$ ), such that the fourth derivative is negative. We then calculate the temperature, using the relation $\beta = \frac{\partial s}{\partial \epsilon}$, 

\begin{equation}
\beta =\mp \frac{1}{K \gamma ^{1/2}} ln ~ \bigg\{\frac{2(1-q_{\pm})}{\sqrt{q_{\pm}^2-m^2}}\bigg\}
\label{eq:T}
\end{equation}
which gives the equivalent phase diagram in the ($T-\Delta-H$) space.

\subsection{Repulsive Blume-Emery-Griffiths Model} \label{sec:mc_a}
\begin{figure}
\centering
\includegraphics[scale=0.7]{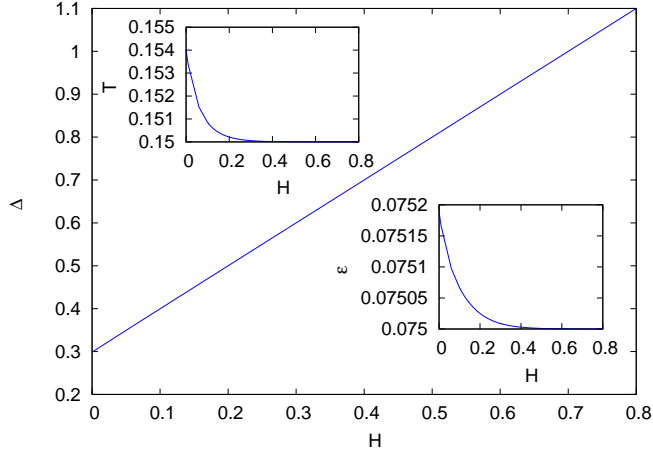}
\caption{The value of crystal field($\Delta$) and temperature($T$) along the $\lambda_+$ line at $K=-0.4$ in the microcanonical ensemble. The main plot shows the $\lambda_+$ line in the  $\Delta-H$ plane, where the value of $\Delta$ increases almost linearly with $H$. From our numerical data, the variation of this line comes out to be $\Delta \simeq (K+1)/2 + H$. \textbf{Bottom Inset}: The $\lambda_+$ line in the $\epsilon -H$ plane. The value of $\epsilon$ decreases and finally saturates at $\epsilon_{sat}$, which is numerically predicted to be $(K + 1)/8$.  \textbf{Top Inset}: The $\lambda_+$ line in the $T-H$ plane, showing similar qualitative behaviour as in the $\epsilon-H$ plot. The value of $T$ saturates for large $H$ at $(K + 1)/4$.}
\label{fig:MC_wings}
\end{figure}

In this section, we show our results for repulsive BEG model in the microcanonical ensemble in the $(T-\Delta-H)$ space. In the absence of magnetic field, this model has been recently studied in \cite{hovhannisyan2017complete,prasad2019ensemble}. We find that for $-0.0828 \leq K \leq 0$ the phase diagram consists of a TCP where the $\lambda_{\pm}$ lines meet the $\lambda$ line in the $H=0$ plane. As the $K$ decreases further,  for $ -1 < K < -0.0828$, it was reported earlier in \cite{prasad2019ensemble} that a critical point (CP) appears in the ordered region of the system along with a CEP. In this topology, as we switch on the field $H$, we note that the $\lambda_{\pm}$ lines meet at the proposed CP. Thus, the CP is actually a BEP. We show our results for $K=-0.4$ in the ($\Delta,H$), ($\epsilon ,H$) and ($T,H$) plane in Fig.\ref{fig:MC_wings}. Here, we show the behaviour of the $\lambda_+$ line for positive $H$. We note that the value of $\Delta$ on the $\lambda_+$ line increases with $H$ almost linearly in the large $H$ limit. The values of $\epsilon$ and the $T$ decreases with $H$ and saturates for large $H$.  We note that the variation of $\Delta$ in the large $H$ limit is of the type, $\Delta \simeq (K+1)/2 + H$. Also, the saturation values are, $\epsilon_{sat} \simeq (K+1)/8$ and $T_{sat} \simeq (K+1)/4$. The values of ($\Delta$, $\epsilon$, $T$) for BEP and the saturation values of $\epsilon$ and $T$ are listed in Table \ref{tab:MC}.

On the $\lambda_{\pm}$ line, the variation of $\Delta$ and $\epsilon$ (or $T$)  in the limit $H \rightarrow \infty$ can be explained in a simple way. In the limit $H \rightarrow \infty$, we can safely assume that there are no particles with spin $-1$, or in other words, $N_-=  0$. Thus, $q$ will be equal to $m$. In this limit, the entropy of the system (per particle) can be written as, $s = - (1-m) ln(1-m) - m ln(m)$, having a maximum at $m=1/2$. Now, the energy per particle, in this limit, turns out to be, $\epsilon \rightarrow \{(\Delta -H)/2 - (K + 1)/8\}$. In order for the energy (per particle) to be finite on the transition line, $\Delta$ should also increase linearly with $H$. We indeed get the linear variation of $\Delta$ with $H$ on the $\lambda_{\pm}$ line. If we use the variation of $\Delta$ as approximated numerically, i.e., $\Delta \simeq (K+1)/2 + H$, we can estimate the saturation value of $\epsilon \rightarrow \epsilon_{sat} \simeq (K + 1)/8$. Using these values in the expression for calculating the temperature (Eq.(\ref{eq:T})), it can be easily shown that the saturation of $T$ will be $T_{sat} \simeq (K+1)/4$. Hence, the saturation values of $\epsilon$ and $T$ will become zero for $K=-1$.

\begin{table*}
\begin{center}
\begin{tabular}{|c|c|c|c|c|c|c|c|}
\hline
\multicolumn{8}{|c|}{Microcanonical: $-1 \leq K \leq 0$}\\
\hline
$K$ & \multicolumn{3}{|c|}{TCP / BEP}  & $\epsilon_{sat}$ & $T_{sat}$ & $\epsilon_w$ & $T_w$\\
\hline
 & $\Delta$ & $\epsilon$ & $T$ & $\simeq (K + 1)/8$& $\simeq (K+1)/4$ & (= $\epsilon_{TCP/BEP} - \epsilon_{sat})$ &  (= $T_{TCP/BEP} - T_{sat})$\\
\hline
0 &0.46240 &0.15275 &0.33033    &0.12502   & 0.25007 & 0.02773 & 0.08026\\
\hline
-0.05 &0.44741 &0.14125 &0.31032  &0.11875   & 0.23750 & 0.0225 & 0.07282\\
\hline
-0.1 &0.43079 &0.12556 &0.27964  &0.11250   & 0.22500 & 0.01306 & 0.05464\\
\hline
-0.2 &0.39100 &0.10343 &0.22454  &0.10000   & 0.20000 & 0.00343 & 0.02454\\
\hline
-0.3 &0.34624 &0.08837 &0.18564  &0.08750   & 0.17500 & 0.00087 & 0.01064\\
\hline
-0.4 &0.29871 &0.07519 &0.15401  &0.07500   & 0.15000 & 0.00019 & 0.00401\\
\hline
-0.5 &0.24968 &0.062529 &0.126145  &0.06250   & 0.12500 & 0.000029 & 0.001145\\
\hline
-0.6 &0.19996 &0.0500025 &0.1001879  &0.05000000   & 0.1000000 & 0.0000025 & 0.0001879\\
\hline
-0.7 &0.15 &0.0375000615 &0.075009205  &00.03750000  & 0.0750000 & 0.000000061 & 0.0000092\\
\hline
-0.75 &0.125 &0.0312500035 &0.06250078  &0.031250000  & 0.0625000 & 0.0000000035 & 0.00000078\\
\hline
-0.8 &$\simeq 0.1$ &$\simeq 0.02500000103$ &$\simeq 0.0500000192$  &$\simeq 0.025$ & $\simeq 0.05$ & $\simeq 0.00000000103$ & $\simeq 0.0000000192$\\
\hline
\end{tabular}
\end{center}
\caption{Co-ordinates of the multicritical points (TCP, BEP), saturation values of $\epsilon$,  $T$ and the width of the wings for $-1 < K \leq 0$.}
\label{tab:MC}
\end{table*}

\begin{figure}
\centering
\includegraphics[scale=1.4]{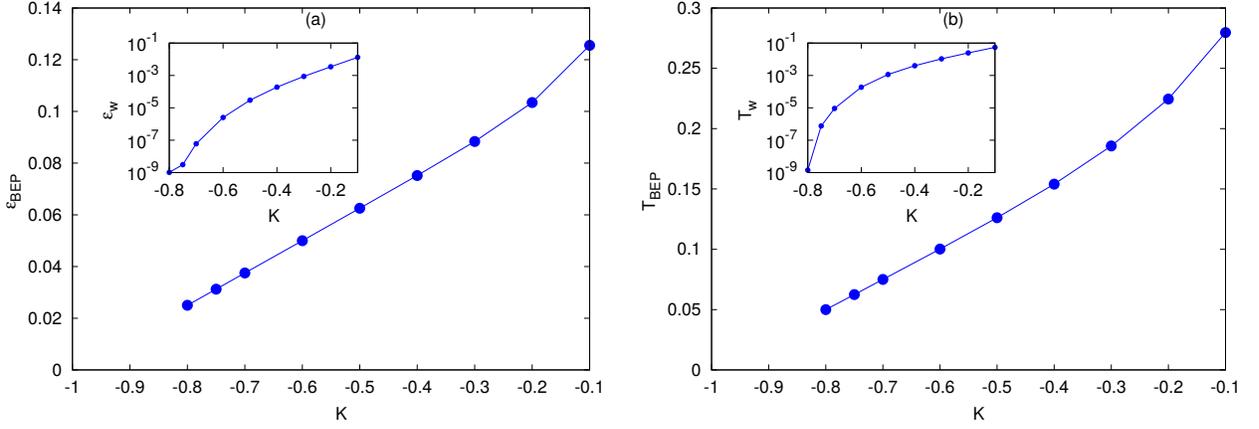}
\caption{Variation of BEP and the width of the wings ($\epsilon_w$ and $T_w$) with $K$. (a) $\epsilon_{BEP}$ decreases as $K$ tends to $-1$, and appears to meet at $\epsilon=0$ at $K=-1$. Inset show the variation of the width of wings in $\epsilon$. We note that the width decreases exponentially as $K$ tends to $-1$. (b) $T_{BEP}$ with $K$ showing similar qualitative behaviour as $\epsilon_{BEP}$. Inset show the width of the wings in temperature $T_w$, which also decreases exponentially as $K$ tends to $-1$.}
\label{fig:MC_BEP}
\end{figure}

We measure the width of the wings in energy(and temperature). We denote it by $\epsilon_w$ (and $T_w)$. We also list the saturation values of $\epsilon$ and $T$ and the width of the wings ($\epsilon_w$ and $T_w$) in Table \ref{tab:MC}. We plot the BEP and the width of the wings ($\epsilon_w$ and $T_w$) in Fig.\ref{fig:MC_BEP}. We note that the BEP tends to $\epsilon=T=0$ as $K \rightarrow -1$. The width of the wings are also found to decrease exponentially and tends to zero as $K$ tends to $-1$. From all the above observations, it is clear that at $K=-1$ the width of the wings vanish and BEP reaches $\epsilon=T=0$. Thus, for $K \leq -1$, there is no phase transition in the non-zero $H$ plane for a finite $T$.


\subsection{Attractive Blume-Emery-Griffiths model } \label{sec:mc_b}
The attractive BEG model has been studied earlier in microcanonical ensemble in \cite{hovhannisyan2017complete}, in the ($T-\Delta$) plane. The full phase diagram($T-\Delta-H$) was not studied before for the microcanonical ensemble as of our knowledge. In this section, we present results for the attractive BEG model in the ($T-\Delta-H$) space. We find that, in the range $0 < K < 3$, the phase diagram is similar to the case $0 > K \geq -0.0828$. For $K > 3$, the $\lambda$ line truncates on the first order line at a CEP. The first order line continues to exist in the paramagnetic region and becomes a surface in the ($T-\Delta-H$) space which separates
two paramagnetic phases P1 and P2(discussed before in Sec.\ref{attrac_beg_cano}) and the wings no longer exist.

\begin{figure}
\centering
\includegraphics[scale=1.4]{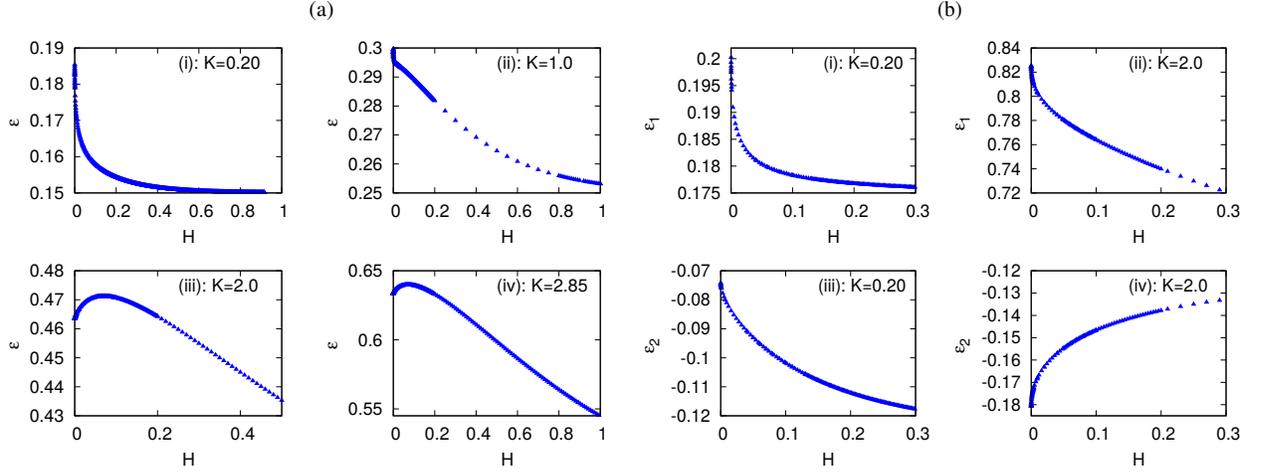}
\caption{Non-monotonic variation of the $\epsilon$ as a function of $H$ along the $\lambda_+$ line for positive $K$. (a) The variation of $\epsilon$ along the $\lambda_+$ line for various $K$. For small $K$, the curve is monotonic; $\epsilon$ decreases with $H$ and then saturates. For large $K$, $\epsilon$ varies non-monotonically with $H$. (b) Variation of $\epsilon_1$ and $\epsilon_2$ for $K=0.20$ and $K=2.0$. The variation of $\epsilon_1$ is similar for small and large $K$ values, however, the qualitative nature in the variation of $\epsilon_2$ is different for small and large $K$. This is the cause of the non-monotonic variation in $\epsilon$.}
\label{fig:positive_K}
\end{figure}

For small positive $K$, the variation of $\epsilon$ is monotonic with $H$ on the $\lambda_{\pm}$ lines, similar to negative $K$. For large positive $K$($\geq 1$), however, the variation in $\epsilon$ is non-monotonic on the transition line as shown in Fig.\ref{fig:positive_K}(a). This can be understood by separating the expression of $\epsilon$ into two parts: $\epsilon =  \epsilon_1 + \epsilon_2$, where, $\epsilon_1 = \Delta q - \frac{1}{2} m^2 -H m$ and $\epsilon_2 = - \frac{K}{2} q^2 $. We note that the variation of $\epsilon_1$ remains similar for small as well as large $K$, however, the variation of $\epsilon_2$ is different for small and large $K$. It decreases with $H$ for small $K$ while increases with $H$ for large $K$ (see Fig.\ref{fig:positive_K}(b)). The variation in $\epsilon_2$ is mainly due to the variable $q$, which itself shows such behaviour. In $\epsilon_1$ also, we have the variable $q$, but it appears with other terms. $\epsilon_1$ does not change its qualitative behaviour when we change $K$. For small $K$, since both the $\epsilon_1$ and $\epsilon_2$ decreases with $H$, the sum also decreases with $H$. For large $K$, there is a competition between $\epsilon_1$ and $\epsilon_2$. In the small $H$ regime, the variation in $\epsilon_2$ dominates, which gives rise to an increase in $\epsilon$ with $H$. For large $H$, the variation in $\epsilon_1$ starts dominating and $\epsilon$ decreases with $H$. For very large $H$, $\epsilon_1$, $\epsilon_2$ and $\epsilon$ will all finally saturate. The saturation values of $\epsilon$ follow similar relationship with $K$ as obtained for the negative $K$. The variation of $T$ also shows similar non-monotonic behaviour in the same range of $K$. We made similar observations in the canonical ensemble in Sec.\ref{attrac_beg_cano}.


\section{Ensemble inequivalence} \label{sec5}
The inequivalence of different ensembles in the Blume-Emery-Griffiths model  has been reported earlier in \cite{hovhannisyan2017complete,prasad2019ensemble} in the absence of magnetic field. In the $(T-\Delta)$ plane, while the $\lambda$ line equation is same in both the ensembles, the first order line and the multicritical points are known to be located differently \cite{mukamel2012,mukamel2010,
mukamel2001,mukamel2014}. It was reported that the TCP and other multicritical points are different for canonical and microcanonical ensembles for a given value of $K$ \cite{hovhannisyan2017complete,prasad2019ensemble}. 

\begin{figure}
\centering
\includegraphics[scale=0.7]{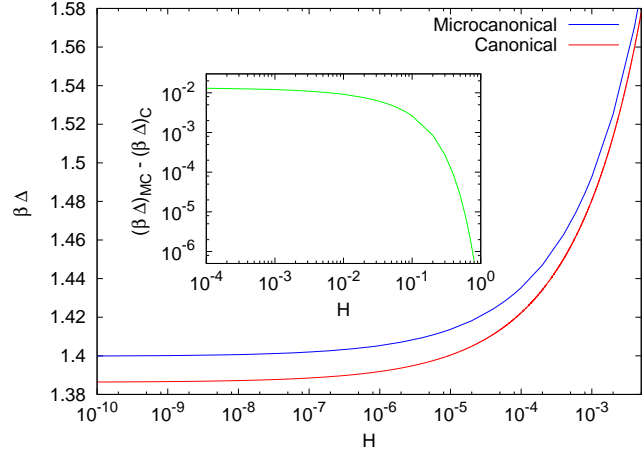}
\caption{Ensemble inequivalence in the Blume-Capel model ($K=0$). We show the locus of the $\lambda_+$ line (product of $\beta \Delta$) as a function of $H$, which is different for the two ensembles. In the inset, we plot the difference in the value of $\beta \Delta$ for the two ensembles, as a function of $H$. This value decreases to zero almost exponentially.}
\label{fig:inequivalence_BC}
\end{figure}

In this work, we have looked at all the three continuous transition lines($\lambda,\lambda_+,\lambda_-$) and the first order surfaces. We find that not just the multicritical points, the continuous transition lines $\lambda_+$ and $\lambda_-$ are also different in the two ensembles. In fact, the ensemble inequivalence of the two ensembles can be seen as a consequence of this inequivalence. For $K=0$, which corresponds to the Blume-Capel model, in Fig.\ref{fig:inequivalence_BC}, we plot the locus of the $\lambda_+$ line in two ensembles and one can see that they are different ($\lambda_-$ line also behaves in a similar way). We plot the product of $\beta \Delta$ on the $\lambda_+$ line as a function of $H$ for both the ensembles, and note that for $H \rightarrow 0$, these lines meet at different points, which is the TCP of their corresponding ensembles. For canonical ensemble, these $\lambda_\pm$ lines meet at $(\beta \Delta)_{TCP} \simeq 1.3863$, while for microcanonical ensemble, these lines meet at $ (\beta \Delta)_{TCP}\simeq 1.3998$ (see Fig.\ref{fig:inequivalence_BC}). We also note that the $\lambda_+$ lines for the two ensembles become close to each other for large $H$. We plot the difference in the value of $\beta \Delta$ for the two ensembles for a given $K$, and plot it as a function of $H$ in the inset of Fig.\ref{fig:inequivalence_BC}. We note that this value decreases exponentially to zero as $H$ becomes large.

\begin{figure}
\centering
\includegraphics[scale=1.4]{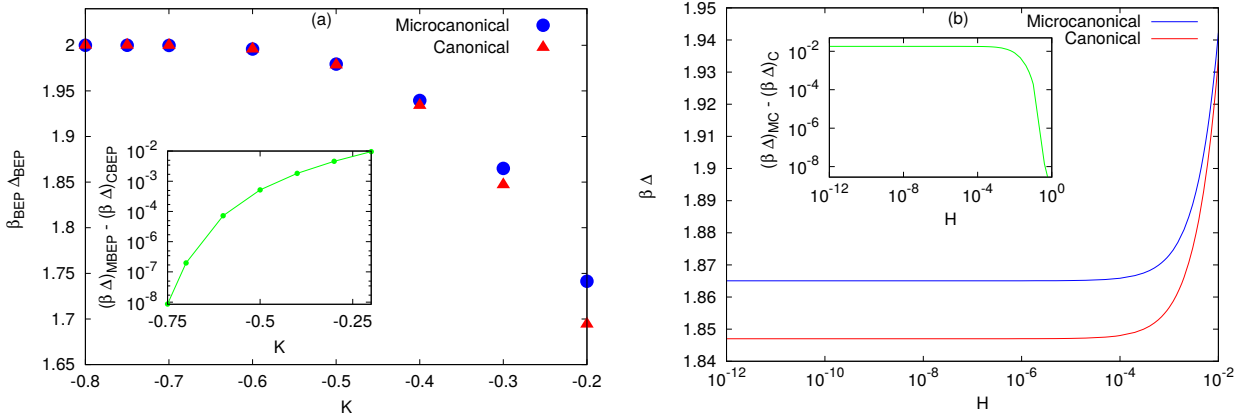}
\caption{Ensemble inequivalence in the Blume-Emery-Griffiths model ($K \neq 0$). (a) The product of $\beta \Delta$ at the BEP, as a function of $K$. We note that the difference in the $\beta _{BEP} \Delta_{BEP}$ decreases to zero as $K \rightarrow -1$. (b) The locus of $\lambda_+$ line (product of $\beta \Delta$) for $K=-0.3$, for the two ensembles. These lines are different in the two ensembles in the small $H$ regime, however, the lines tend to become closer as $H$ increases. In the inset, we plot the difference in the value of $\beta \Delta$ as a function of $H$, for the two ensembles. The difference decreases with increasing $H$.}
\label{fig:inequivalence}
\end{figure}

For non-zero $K$, the $\lambda_\pm$ lines meet the $\lambda$ line at the TCP. This topology persists for $0 \geq K \geq -0.1838$ in the canonical ensemble, whereas for microcanonical ensemble this topology occurs for $0 \geq K \geq -0.0828$. As $K$ decreases further(for canonical ensemble $-0.1838 < K < -1$ and for microcanonical ensemble $-0.0828 < K < -1$), the $\lambda_\pm$ lines move inside the ordered region and meet at BEP in the $H=0$ plane. Interestingly, we find that the difference in the position of BEP and CEP in the two ensembles decreases with decreasing $K$ and for $K=-1$ the two ensembles become equivalent. In Fig.\ref{fig:inequivalence}(a), we plot the value of $\beta \Delta$ at the BEP for both the ensembles. We note that the value of $\beta_{BEP} \Delta_{BEP}$ for the two ensembles becomes closer as $K \rightarrow -1$. In the inset of Fig.\ref{fig:inequivalence}(a), we also plot the difference in the value of $\beta_{BEP} \Delta_{BEP}$ for microcanonical and canonical ensembles, and note that this difference decreases exponentially as $K \rightarrow -1$. Thus, for $K \leq -1$, we find that there is no ensemble inequivalence in the $H=0$ plane.

We have shown in Sec.\ref{rep_beg_can} and Sec.\ref{sec:mc_a} that for $K \le -1$, there is no phase transition for finite magnetic field in either of the ensembles and hence there are no wings. Thus there is no inequivalence in the $H \neq 0$ plane as well. For $K > -1$, however, we do have wings and the continuous transition lines $\lambda_+$ and $\lambda_-$, meet at its corresponding TCP or BEP for canonical and microcanonical ensembles in the limit $H \rightarrow 0$. Thus, the critical lines in the $H$ plane are different for the two ensembles for $K > -1$. In Fig.\ref{fig:inequivalence}(b), we plot the value of $\beta \Delta$ on the continuous transition line for $K = -0.3$, as a function of $H$ for both the ensembles. We note that the two lines are different for small $H$, however, these lines tend to meet each other for large $H$. We measure the difference between the value of $\beta \Delta$ for the two ensembles for a given $K$, and plot it as a function of $H$  in Fig.\ref{fig:inequivalence}(b) inset. We note that this difference reaches zero almost exponentially as $H$ increases. Thus, in the limit $H \rightarrow \infty$, these critical lines for both the ensembles become equivalent.

From the above discussion, it is clear that the ensemble inequivalence is observed for $K > -1$ with small $H$ values, however, for $K \leq -1$, the phase diagrams in the two ensembles become equivalent. In previous literature \cite{mukamel2005,hugo2004,mukamel2010,mukamel2010b}, where ensemble inequivalence with long-range interactions are studied, it was found that whenever the two ensembles (either micro-canonical/canonical or canonical/grand-canonical) have a continuous transition, the transition occurs at the same point. The phase diagrams of the two ensembles can however be different from each other when the phase transition becomes first order in one of the ensembles. This kind of behaviour is observed in many systems such as the spin-1 Blume-Emery-Griffiths (BEG) model \cite{mukamel2005,hugo2004}, the ABC model\cite{mukamel2010,mukamel2010b} etc.

 However, this is not always true. Even the continuous transition point can be different in the two ensembles. For example in a generalized ABC model in Ref. \cite{barton2011}, the canonical and grand-canonical ensembles are found to exhibit a second-order phase transition at different points in the phase space. In \cite{mukamel2012}, a general statement is provided to check the possibility of ensemble inequivalence for continuous transition using Landau theory. The transition is observed for a system undergoing phase transition governed by some order parameter, `$\mu_1$' (say) in a given ensemble. This parameter can be the average magnetization in the case of a magnetic transition, or the difference in the density of the two phases for a liquid-gas phase transition. Then the model is  considered within a `higher' ensemble, where a certain thermodynamic variable, denoted by `$\mu_2$', is allowed to fluctuate, (within the `lower' ensemble, `$\mu_2$' was kept at a fixed value). In the case where $\mu_2$ is the energy, the two ensembles would correspond to the canonical and micro canonical ensembles, while in the case when $\mu_2$ is the particle density, they would correspond to the grand-canonical and canonical ensembles. The system is thus described by the Landau free energy denoted by $f (\mu_1,\mu_2)$. They found that  \cite{mukamel2012} if $f (\mu_1,\mu_2) = f (-\mu_1,\mu_2)$, the two ensembles will be equivalent, when any of them shows a continuous transition. If on the other hand, $f (\mu_1,\mu_2) = f (-\mu_1,-\mu_2)$, the system will show ensemble inequivalence even for continuous transition.

In our case, $\mu_1$ is the magnetization $m$, and $\mu_2$ is the energy $\epsilon$ of the system. The lower ensemble in our case is thus the microcanonical ensemble and the higher one is the canonical. If we check the above symmetries, we note that neither of the conditions studied in \cite{mukamel2012} is satisfied. When we add a magnetic field term, the symmetry of the problem is broken and we find that we have ensemble inequivalence even when the two ensembles show second order transition.


\section{Conclusion} \label{sec6}

The repulsive and attractive BEG model in canonical and in microcanonical ensemble has been studied earlier. This model is known to  exhibit many multicritical points along with the first and second order line of transition. Earlier the model was studied in the ($T-\Delta$) plane\cite{blume1971ising, hovhannisyan2017complete,PhysRevA.11.2079, mukamel1974ising, PhysRevB.15.441, berker1976blume, BAKCHICH1992524, tanaka1985spin, buzano1993surface, chakraborty1986spin, saul1974tricritical, hoston1991multicritical, prasad2019ensemble, hoston1991dimensionality, PhysRevB.47.15019, BRANCO1996477, wang1987phase, wang1987phase1, ekiz2002multicritical,rachadi2004monte, netz1992new, rosengren1993phase} and the ensemble inequivalence was reported\cite{hovhannisyan2017complete,prasad2019ensemble}. The full phase diagram in the $(T-\Delta-H)$ space was studied only for the attractive BEG model in the presence of external field in the canonical ensemble\citep{mukamel1974ising}. We revisited the model in order to study the full phase diagram in the ($T-\Delta-H$) space in both the ensembles on a complete graph. Though we  explored the phase diagram for the entire range of $K$, we mainly focused on the repulsive BEG model. We found that for small negative $K$, the model exhibits a TCP where the $\lambda_{\pm}$ lines meet. For the canonical ensemble the range of $K$ for such a topology was $0 \geq K \geq -0.1838$, whereas for microcanonical ensemble it was $0 \geq K \geq -0.0828$. As $K$ decreases further, the wings meet inside the ordered phase at a BEP(for canonical $-0.1838> K > -1$ and for microcanonical $-0.0828> K > -1$). This point was identified as an ordered critical point 
in the earlier studies \cite{prasad2019ensemble}. We also observed that as $K \rightarrow -1$, the width of the wings decreases. At exactly $K=-1$, the wing width in temperature becomes zero along with the CEP and BEP reaching $T=\Delta=0$ in both canonical and microcanonical ensemble. For $K> -1$, we observe that the $\lambda_\pm$ lines are different in the two ensembles and they meet at different multicritical points in the $H=0$ plane. For $K \leq -1$, we find that there is no phase transition in the $H$ plane for both the ensembles. 

Absence of transition in the $H$ plane for $K \le -1$ can be argued by looking at the energy. The energy of the system in terms of the order parameters can be written as:  $ \epsilon =- \frac{1}{2} (m^2 +K q^2)+ \Delta q -H m $. For low temperatures we can take $m \approx q$. In that case, only for $K>-1$ the first term can lower the energy and can take over the entropy at sufficiently small temperatures. Hence the transition in $H$ plane is likely only for $K>-1$.

Disorder, in general, is known to smoothen the first order transition and has been known to convert a TCP into a BEP \cite{mukherjee2020emergence}. We studied a pure BEG model here.  We found that the competition induced by negative $K$ affects the phase diagram in a similar manner. It would be interesting to see if there is a similarity in the phase diagram of the two problems even in finite dimensions.

It was shown\cite{PhysRevE.70.046111} in the three dimensional Blume-Capel model by introducing a constraint in the number of vacancies that the presence of the constraint modifies the behaviour near the tricriticality and thus some of the critical exponents get renormalized, although the universality class remains unchanged. This gives rise to a discrepancy between the constraint and the unconstraint system which acts as ensemble inequivalence. It might be also interesting to investigate if such discrepancies exist in the full phase diagram(($T-\Delta-H$) space).

 Also, earlier work on frustrated BEG on  bipartite lattices\citep{hoston1991multicritical,BRANCO1996477} shows that two new ordered phases, namely antiquadrupolar and ferrimagnetic occur for repulsive BEG. These phases are not possible on a complete graph. Hence it would be interesting to study the effect of large negative $K$ on these two phases especially in the mean field limit on a bipartite lattice.


\section{Acknowledgement}

Raj Kumar Sadhu acknowledges National Institute of Science Education Research, Bhubaneswar(INDIA) for funding the visiting research program during the initiation of this project.

\appendix
\renewcommand{\theequation}{A-\arabic{equation}}
\setcounter{equation}{0}
\renewcommand{\thefigure}{A-\arabic{figure}}
\setcounter{figure}{0}


\section{Calculation of Free energy functional} \label{app2}

In order to solve the Hamiltonian(Eq.(\ref{eq:h})), we take the non-interacting Hamiltonian: $H\sum_{i} S_i-\bigtriangleup \sum_{i} S_i^2 $ , with the probability measure:

\begin{eqnarray}
P(1)=\frac{e^{\beta (H- \Delta)}}{1+ 2 e^{- \beta \Delta} \cosh {\beta H})}\\
P(-1)=\frac{e^{-\beta (H+ \Delta)}}{1+ 2 e^{- \beta \Delta} \cosh {\beta H})}\\
P(0)=\frac{1}{1+ 2 e^{- \beta \Delta} \cosh {\beta H})}\\
\end{eqnarray}
The scaled Cumulant generating function(CGF) is:
\begin{eqnarray}
\Lambda(k_1,k_2)&=& \lim_{N \rightarrow \infty} \frac{1}{N} \log {<e^{ N(x_1 k_1 + x_2 k_2)}>_P}\nonumber \\
&=&  \log (1+2 e^{k_2-\beta \bigtriangleup} \cosh (k_1+ \beta H)) -\log (1+2e^{-\beta \bigtriangleup} \cosh \beta H)
\end{eqnarray}
The rate function $R$ for the non interacting Hamiltonian then can be evaluated using Gartner Ellis theorem\citep{TOUCHETTE20091} and is given by :

\begin{eqnarray}
R(x_1,x_2) &=& \sup_{k_1,k_2} [x_1k_1+x_2k_2-\Lambda(k_1,k_2)]\nonumber \\
&=& \sup_{k_1,k_2} [x_1k_1+x_2k_2- \log (1+2 e^{k_2-\beta \bigtriangleup} \cosh (k_1+ \beta H))]+\log (1+2e^{-\beta \bigtriangleup} \cosh \beta H )]
\end{eqnarray}
Minimizing the above equation w.r.t $k_1$ and $k_2$ gives the following relations:

\begin{eqnarray} \label{eq:t1}
&\Rightarrow & x_1 = \frac{2 e^{k_2^*- \beta \Delta} \sinh k_1^*} {1 + 2 e^{k_2^*-\beta \bigtriangleup} \cosh k_1^* } 
\end{eqnarray}

\begin{eqnarray} \label{eq:t2}
&\Rightarrow & x_1 = \frac{2 e^{k_2^*- \beta \Delta} \cosh k_1^*} {1 + 2 e^{k_2^*-\beta \bigtriangleup} \cosh k_1^* } 
\end{eqnarray}
where $k_1^*$ and $k_2^*$ are the minimums of $k_1$ and $k_2$. This implies:
\begin{equation}
\frac{x_1}{x_2}=\tanh k_1^*
\end{equation}
The interacting part of the Hamiltonian is $- \frac{1}{2} x_1^2-\frac{K}{2} x_2^2$. Now the full rate function of the total Hamiltonian  can be obtained by making use of the tilted LDP. Tilted LDP allows us to generate a new large deviation principle(LDP) from an old LDP by a change of measure.

Let $W_n$ is a sequence of a random variable taking values from $\mathcal{H}$ and a subset $A$ of $\mathcal{H}$. We define the probability measures
\[
Q_{n,\Phi}=\frac{1}{Z_{n}}\int_{A}e^{[n\Phi(x)]}P_{n}\{W_{n}\in dx\}
\]
where $Z_{n}$ denotes the normalizing constant. Here $P_n$ is the probability measure on $\mathcal{H}$ which satisfies LDP with rate function $I$ and $\Phi$ is a continuous function mapping
$\mathcal{H}$ into $\mathcal{R}$ which is bounded from above.
Then according to the tilted LDP, the sequence of probability measures \{$Q_{n,\Phi},$ $n\in N$\}
satisfies LDP on $\mathcal{H}$ with the rate function 
\[
I_{\Phi}(x)=[I(x)-\Phi(x)]-\inf_{y\in\mathcal{H}}\{I(y)-\Phi(y)\}
\]
In our system the rate function $R(x_1, x_2)$ can hence be tilted to give the full rate function to be:

\begin{eqnarray}
I(x_1,x_2) &=& x_1k_1^*+x_2k_2^*-\Lambda(k_1^*,k_2^*)-\frac{\beta x_1^2}{2} - \frac{\beta K x_2^2}{2}   - \inf_{k_1, k_2}[R(k_1,k_2)-\frac{\beta k_1^2}{2} -\frac{\beta K k_2^2}{2}]
\label{eq1}
\end{eqnarray}
Minimization of which w.r.t $x_1$ and $x_2$ gives the following free energy functional:
\begin{eqnarray}
\tilde{f}(m,q) 
 &= & \frac{\beta m^{2}}{2} +\frac{\beta K q^{2}}{2} -\log (1+2 e^{\beta(K q-\Delta)} \cosh \beta (m+H) )+\log (1+2 e^{-\beta \Delta} \cosh \beta H) 
 \label{freeenergy1}
\end{eqnarray}
where the minimums are denoted as: $x_1^* =m$, $x_2^* =q$, which gives the Eq.(\ref{freeenergy}) at the fixed points in Sec.\ref{sec3}.

\bibliography{reference}

\end{document}